\documentclass[prb,aps,superscriptaddress,footinbib
,twocolumn
,longbibliography
]{revtex4-1} 

\usepackage{graphicx}
\usepackage{color,colordvi}
\usepackage{dcolumn}
\usepackage{amsmath}
\usepackage{amssymb}
\usepackage
[disable]      
{todonotes}
\usepackage{float}
\usepackage{epstopdf}
\newcommand{\ts}{\textstyle }

\def\beq{\begin{equation}}
\def\eeq{\end{equation}}

\begin{document}

\title{Zeros of the partition function and phase transition}
\author{Wytse van Dijk}
\email{Electronic mail: vandijk@physics.mcmaster.ca}
\affiliation{Department of Physics and Astronomy, McMaster University,
  Hamilton, ON, Canada L8S 4M1}
\affiliation{Physics Department, Redeemer
  University College, Ancaster, ON, Canada L9K IJ4 }
\author{Calvin Lobo}
\affiliation{Department of Physics and Astronomy, McMaster University,
  Hamilton, ON, Canada L8S 4M1}\author{Allison MacDonald}
\affiliation{Department of Physics, University of Alberta, Edmonton, AB, Canada T6G 2E9} 
\author{Rajat K. Bhaduri}
\email{Electronic mail: bhaduri@mcmaster.ca}
\affiliation{Department of Physics and Astronomy,
McMaster University, Hamilton, ON, Canada L8S 4M1}
\date{\today}

\begin{abstract}

The equation of state of a system at equilibrium may be derived from
the canonical or the grand canonical partition function. The former is a
function of temperature $T$, while the latter also depends on the chemical
potential $\mu$ for diffusive equilibrium. In the literature, often the
variables $\beta=(k_BT)^{-1}$ and fugacity $z=\exp(\beta \mu)$ are used instead. 
For real $\beta$ and $z$, the
partition functions are always positive, being sums of positive
terms. Following Lee, Yang and Fisher, we point out that valuable
information about the system may be gleaned by examining the zeros of
the grand partition function in the complex $z$ plane (real $\beta$),
or of the canonical partition function in the complex $\beta$ plane. In
case there is a phase transition, these zeros close in on the real
axis in the thermodynamic limit. Examples are given from the van der
Waal gas, and from the ideal Bose gas, where we show that even for a finite system
with a small number of particles, the method is useful.
\footnote{Part of this paper is based on the results reported in the unpublished undergraduate theses of Calvin Lobo and Allison MacDonald.}   

\end{abstract}

\maketitle

\section{Introduction}
In a senior level undergraduate or a beginning graduate course,  examples of phase
transition are often given from the classical van der Waals equation of 
state, and at a quantum level, from the Bose-Einstein condensation
(BEC) of an ideal Bose gas.\cite{landau58,huang65} 
The treatment, naturally, focuses on the 
equation of state of the system as a function of  physical 
parameters  like the temperature $T$ and chemical potential $\mu$, which are
real.   It is instructive to learn, however, that the approach
towards  phase transition may be studied by examining the analytical behavior 
of the partition function for complex values of the parameters, even 
for a finite system where there is no discontinuity in the
derivatives of the free energy.      

For a system in thermal and diffusive equilibrium, it is convenient to 
calculate the ensemble average using the grand canonical partition
function ${\cal Z}(\beta, z)$, where $\beta =(k_BT)^{-1}$,  $k_B$
being the Boltzmann constant and $z=\exp(\beta\mu)$  the fugacity.
Note that there is an implicit volume dependence in ${\cal Z}$, since
the eigenenergies are volume dependent. We suppress this in our
notation ${\cal Z}(\beta, z)$ for simplicity.  
The grand canonical partition function  is a sum of positive definite
 terms for real positive values of 
$\beta$ and $z$, and as such cannot have a zero in the physical 
domain of these variables.
Lee and Yang\cite{yang52,lee52a} considered a lattice gas with a
hard core interaction.  
Because of the short-range repulsion between the particles, only a
finite number of particles may be packed into a finite volume. As we
shall see, this allows one to express ${\cal Z}$ 
 as a finite-degree polynomial in fugacity $z$. This
polynomial is then completely defined in terms of  its zeros on the
complex fugacity plane. 
These zeros are all complex, coming in 
 complex conjugate pairs. In the thermodynamic limit, the zeros 
  coalesce in continuous lines, tending to pinch the  positive real
 $z$ axis at a phase transition.
 In this paper, we show that this tendency sets in even at finite particle
number and volume, with the zeros moving closer to the real axis as the particle 
number is increased. Even though complex, the closer a zero  comes to the
real axis, the more it dominates real thermodynamic properties.   
 Note that the validity of  the Lee-Yang 
method rests on the repulsive core between the particles.   

Fisher~\cite{fisher65} pointed out that the zeros of the canonical
 partition function $Z_N(\beta)$ on the complex $\beta$ plane have an
 analogous behavior. However, Fisher zeros can give useful information
 even in the absence of the repulsive core. 
The Fisher zeros for ideal trapped bosons were
 studied  in the context of BEC by M\"ulken
{\it et al.}\cite{mulken01}
In this paper, following the work of Hemmer {\it et
  al.},\cite{hemmer66} we study the Lee-Yang zeros of the 
classical van der Waals gas
(that has a phase transition with a critical temperature) and compare 
it with a Calogero gas\cite{calogero69,*calogero69a,sutherland71,*sutherland71a}  (that has no phase transition). 
The ideal Bose gas (which undergoes a phase transition at BEC) 
is studied in some detail, both using  the grand canonical and the canonical formalisms.
The heat capacity per particle in the grand canonical and canonical ensembles 
 are compared at BEC to check how close these are for finite particle number. 
Since there is no short-range repulsion in the ideal Bose gas, the
Lee-Yang zeros are not meaningful, but the Fisher zeros are.
Accordingly, we find the pattern of these on the complex $\beta$ plane
for 50 and 100 atoms. Even for such small number of particles,
we see a clear tendency for the zeros to close in on the real $\beta$
axis. The calculations are done for the exact $Z_N(\beta)$ as well as
the more commonly used continuous density of states.  This will be
discussed  after the patterns of zeros are presented. 
The grand partition function, on the other hand, 
is shown to have a pole  at phase transition  
for real $z=1$.~\cite{ikeda82,wang99} 
A nontrivial modification
over the ideal gas  will be to introduce  interparticle interaction  
through virial coefficients in the grand
potential,\cite{bhaduri12} and study how the 
 zeros of the grand canonical partition function shift on the complex plane.  
This is beyond the scope of the present paper. 

\section{Partition Function}
The canonical partition function (for fixed $N$) is defined  as 
\beq
Z_N(\beta)=\sum_{E_i^{(N)}} \exp(-\beta E_i^{(N)})~,
\label{eq:01}
\eeq
where  $E_i^{(N)}$ are the complete set of
eigenenergies of the $N$-body system including states in the continuum, if any. The sum is taken over all states
${i}$ including the degeneracies. Since the energies $E_i^{(N)}$
depend on the volume of the system, there is a  volume 
dependence in $Z_N(\beta)$.  
The grand canonical partition function, on the other hand, allows for 
particle exchange (in addition to energy) via the reservoir, and is defined as 
\beq
{\cal Z}(\beta, z)=\sum_{N=0}^{\infty}\sum_{E_i^{(N)}}
 \exp(-\beta E_i^{(N)}+\beta \mu N)~=\sum_{N=0}^{\infty}
Z_N(\beta) z^N~,
\label{eq:02}
\eeq
where the fugacity $z=\exp (\beta \mu)$. The simplest example is an
ideal classical gas in a volume $V$.  The $N$-particle canonical
partition function is the $N$th power
of the one-particle  partition function $Z_1(\beta)$. The latter is
calculated by integrating $\exp(-p^2/2m)$ over the phase space divided
by $h^3$, where $h$ is the Planck's constant. The net result is  
\beq
Z_N(\beta)=\frac{1}{N!}~\left(\frac{V}{\lambda_T^3}\right)^N~,
\label{eq:03}
\eeq
where $\displaystyle \lambda_T=\sqrt{\frac{2\pi\hbar^2}{Mk_bT}}$ is the thermal wave
length, and the customary division by $N!$ has been made to preserve
the extensive property of the entropy. By substituting
Eq.~(\ref{eq:03}) in Eq.~(\ref{eq:02}), we obtain 
\beq
\frac{1}{V} \ln {\cal Z}^{(0)}(\beta, z)=\frac{z}{\lambda_T^3}~,
\label{eq:04}
\eeq  
where the superscript on ${\cal Z}$ denotes a non-interacting system. 

More generally, for an interacting gas with short-range interparticle
repulsion, 
Eq.~(\ref{eq:02}) shows that ${\cal Z}$ is a {\it finite degree} polynomial
in $z$, and therefore may be completely defined in terms of its
zeros. 
These zeros, however, cannot be on the
real  positive $z$-axis, since every term in Eq.~(\ref{eq:02}) is then
positive. Accordingly, for complex $z$ (but real 
positive $\beta$), Eq.~(\ref{eq:04}) may be generalized to~\cite{fisher65}
\beq
{\cal Z}(\beta, z)= \prod_r\left(1-\frac{z}{z_r}\right)~.
\label{eq:05}
\eeq
 The zeros  $z_r(\beta, V)$ come in complex conjugate pairs since the
coefficients $Z_N(\beta)$ of the polynomial (\ref{eq:02}) are real. 
In case there is a
phase transition at some temperature, a zero and its complex conjugate 
tend to pinch the real $z$ axis. If there are more than one phase transitions,
there are segments on the real positive $z$-axis  that are zero-free. 
 The grand potential $\Omega=-k_BT \ln {\cal
  Z}(\beta,z)=-PV$, hence $\displaystyle\beta P=\frac{1}{V} \ln {\cal  Z}(z,\beta)$.  
Yang and Lee~\cite{yang52} proved in general that for 
$V\rightarrow\infty$, this limit exists, and $P$ increases monotonically
in the zero-free segments. 
At the interface of two phases, the pressure $P$ remains continuous,
but its slope as a function of $\ln z$ is not the same. Moreover, 
in the thermodynamic limit, the number density 
$\rho=\frac{\ts\partial}{\ts\partial \ln z}(\beta P)$ may be discontinuous as a
function of $\ln z$ in the interface of two phases (see Finkelstein.\cite{*[{}] [{, chapter 10.}]  finkelstein69})

 For our example in the next
section,  it is more relevant to take a finite volume $V$ which
contains $N$ particles. 
 If there is a hard-core repulsion, then there is a maximum
number $N_\mathrm{max} $ that can be accommodated in this volume. Then the
infinite upper limit in the sum over $N$ in Eq.~(\ref{eq:02}) is
replaced by $N_\mathrm{max}$, and 
 ${\cal Z}(\beta, z)$ is a polynomial of order $N_\mathrm{max}$.

\subsection{van der Waals gas}
 This is a classical example, studied in the context of Yang-Lee zeros
 by Hemmer {\it et al.}~\cite{hemmer66}   Equation~(\ref{eq:02}) is used with the canonical partition
 function that is postulated to be  
\beq
Z_N(\beta)=\frac{1}{N!} (V-N b)^N~\exp \left(\frac{2 a \beta}{V} 
\frac{N(N-1)}{2}\right)~,      
\label {eq:07}
\eeq  
where $b$ is interpreted as the volume associated with the repulsive
core,  and $a$ as a measure of the outer attraction. 
For calculating ${\cal Z}(\beta, z) $ using Eq.~(\ref{eq:02}), one takes
$N_\mathrm{max}=V/b$. Note that the excluded volume effect, and the outer 
pair-wise attraction are both incorporated in the canonical
$Z_N(\beta) $ above. The equation of state can easily be deduced from 
the postulated $Z_N(\beta)$. One obtains the Helmholtz free
energy $F=-1/\beta \ln Z_N(\beta)$, and the pressure
$\displaystyle P=-\left(\frac{\partial F}{\partial V}\right)_T$. A little algebra then yields
the equation of state 
\beq
\left (P+\frac{N^2}{V^2}a\right )(V-Nb)=Nk_B T~.
\label{eq:08}
\eeq   
where we have assumed $N\gg 1$. One makes the Maxwell construction 
across the unphysical region in which $P$ decreases with $V$ to obtain 
the equation of state. The critical point is obtained by additionally imposing the
condition that the first and second partial derivatives of $P$ with
respect to $V$ (at constant temperature ) are zero; see for example 
Landau and Lifshitz.\cite{landau58} The critical temperature is given by 
\begin{figure*}[t]
\centering                                                                                      
$\begin{array}{c@{\hspace{0.0in}}c@{\hspace{0.0in}}c}                                                                                                             
\resizebox{2.3in}{!}{\includegraphics[angle=270]{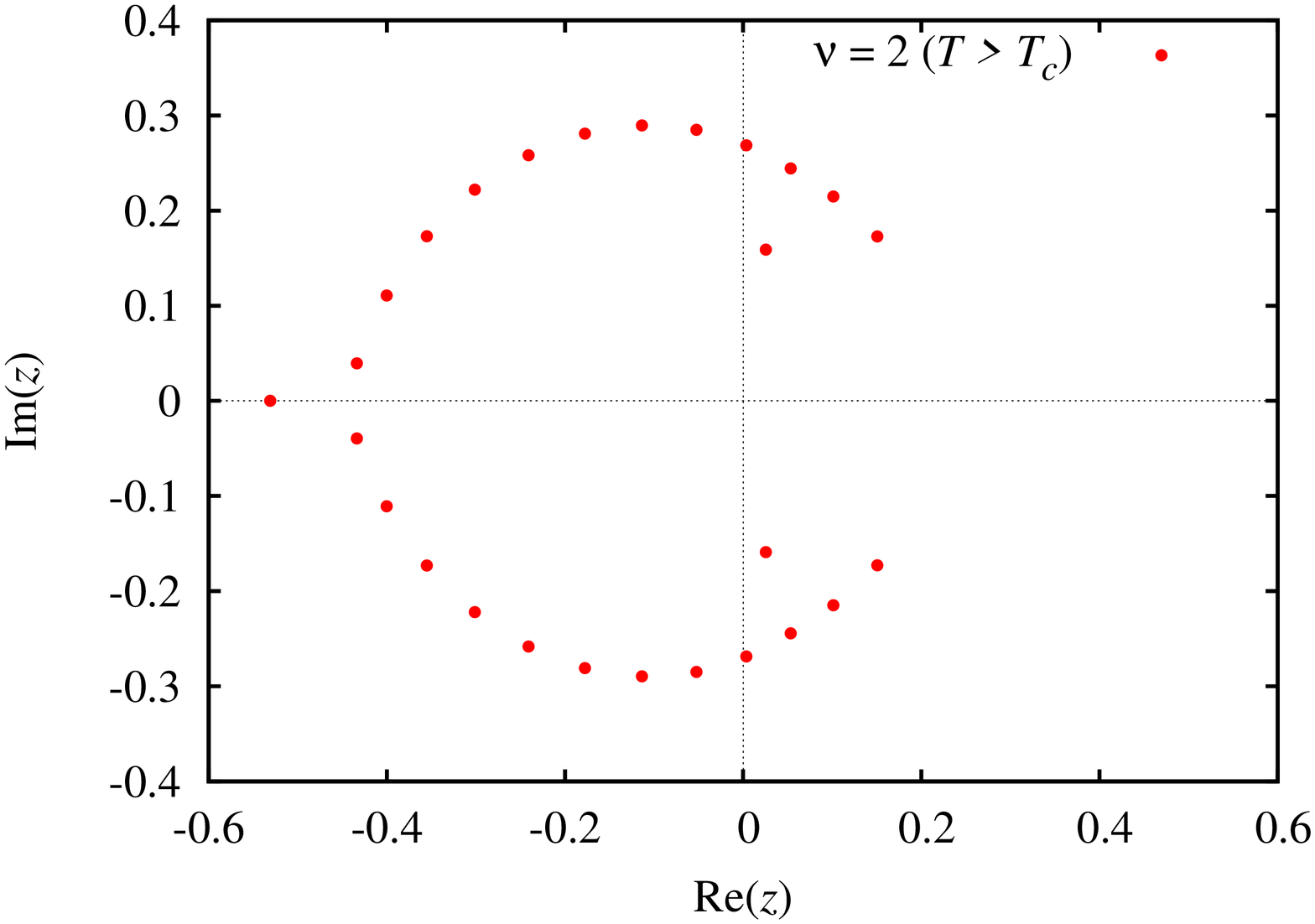}} &                                    
\resizebox{2.3in}{!}{\includegraphics[angle=270]{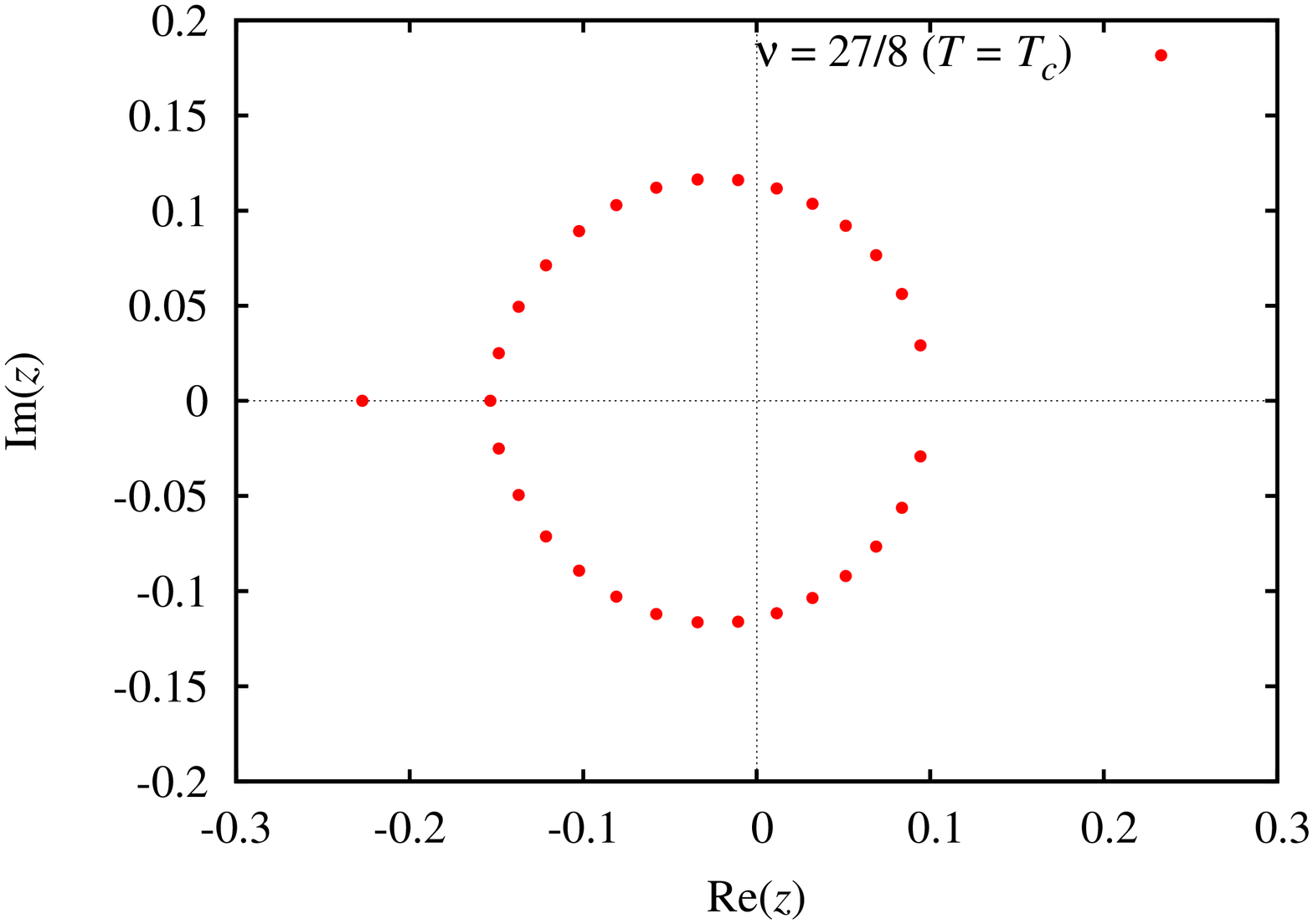}} &                                  
\resizebox{2.3in}{!}{\includegraphics[angle=270]{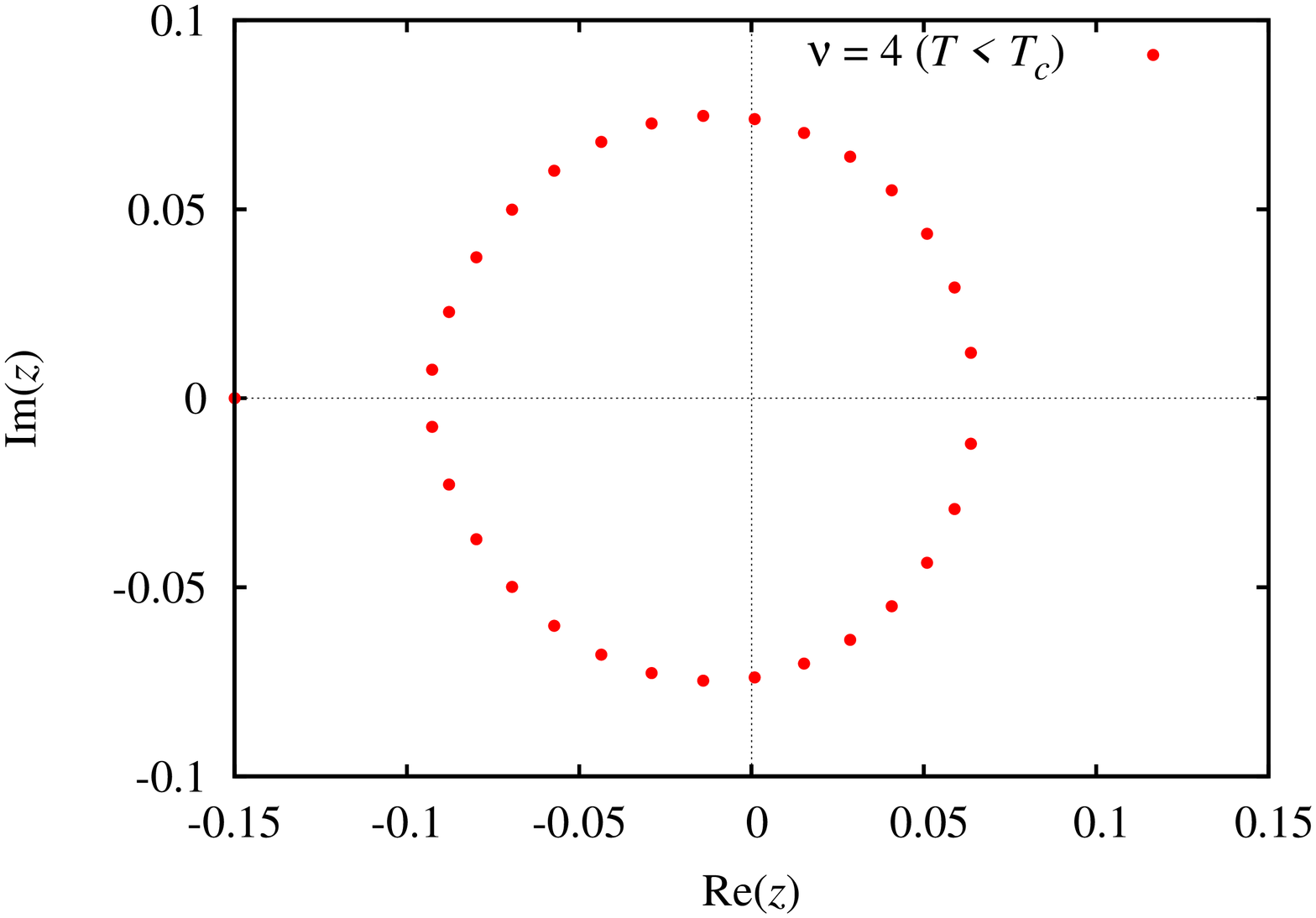}}                                      \end{array}$                                                                                                                                                                                
\caption{The Yang-Lee zeros of the grand partition function for a van der Waals                     
 system of 40 particles, i.e., $V=40$, $d=1$, $N=40$.  From left to right the patterns              
correspond to $\nu=2$, $\nu=27/8$, and $\nu=4$. }                      
\label{fig:01}                                                                             
\end{figure*}
\beq
k_BT_c=\frac{8}{27} \frac{a}{b}~.
\label{eq:09}
\eeq
  \todo[inline]{(Programs:                                         
  /home/vandijk/2013/paper\_LY\_zeros/van\_der\_Waal/maple/V\_30\_nu\_?.eps.)}
Our interest here is to compute the 
zeros of ${\cal Z}(\beta, z)$ using Eqs.~(\ref{eq:02}) and (\ref{eq:07}) in the 
complex $z$-plane for real values of $T$. From Eq.~(\ref{eq:07}), we see
that there is a cut-off in the upper limit of the summation over $N=
N_\mathrm{max}=V/b$, which prevents us from obtaining ${\cal Z}$ analytically. 
Note, from Eq.~(\ref{eq:09}),
that $a/b$ has the dimension of energy. Setting  $a=\nu k_BT$, and $b=1$,
Eq.~(\ref{eq:09}) takes the form 
\beq
\nu=\frac{27}{8} \frac{T_c}{T}
\label{eq:10}
\eeq
For our calculations, we take $N_\mathrm{max}=V/b=40$. In Fig.~\ref{fig:01}, we plot 
the zeros of ${\cal Z}(T,z)$ for three choices of $\nu$, corresponding
to $T>T_c, T=T_c,$ and $T<T_c$.  Even for 
$N_\mathrm{max}=40$, we clearly see the zeros closing in on the real $z$-axis  
for $T\leq T_c$. 
system of 40 particles, i.e., $V=40$, $d=1$, $N=40$.  From left to right the   \todo[inline]{(Programs:                                         
  /home/vandijk/2013/paper\_LY\_zeros/van\_der\_Waal/maple/V\_30\_nu\_?.eps.)}

\subsection{Calogero gas}
The Calogero gas is an exactly solvable one-dimensional model where point
particles are interacting with a pair-wise inverse-square 
potential.\cite{calogero69,*calogero69a,sutherland71,*sutherland71a} 
The particles are trapped in a harmonic oscillator (HO) potential. For a
repulsive interaction, the high temperature limit of the canonical 
partition function is given by\cite{bhaduri10} ($\hbar=1$) 
\beq
Z_N(\beta)= \frac{1}{N!} \frac{1}{(\beta\omega)^N} \exp\big(-\alpha
\beta\omega N(N-1)/2\big)~,
\label{eq:11}
\eeq
where $\omega$ is the oscillator frequency, and $\alpha$ is a measure 
of the strength of the inverse-square two-body potential. 
Since the density of states is a constant in a one-dimensional HO, it 
is like a two-dimensional gas. The oscillator length is
$l=\sqrt{\hbar/M\omega}$, and we may define a density 
$n=N/l^2=N\omega$, with $\hbar=M=1$. The thermodynamic limit is 
taken as $N\rightarrow \infty, \ \omega\rightarrow 0$, with $N\omega=n$ 
a constant. For $N\gg 1$, we then get
\beq
Z_N(\beta)= \frac{1}{N!} \frac{1}{(\beta\omega)^N} \exp(-\alpha
\beta nN/2)~.
\label{eq:12}
\eeq
With this $Z_N(\beta)$, the grand partition function ${\cal Z}(\beta,
z)$ may be obtained analytically by summing over all $N \rightarrow \infty$ . This is so 
because these are point particles with no excluded volume. A little
algebra immediately gives 
\beq
\ln {\cal Z}(\beta, z)=\frac{z}{\beta\omega} \exp (-n \alpha\beta/2).
\label{eq:13}
\eeq 
This is of the same form as Eq.~(\ref{eq:04}) of the perfect classical
gas, monotonically increasing with $z$. There is, of course, no phase transition.

\section{Ideal Trapped Bosons and BEC}

\begin{figure}[ht]
\centering
\resizebox{3.5in}{!}{\includegraphics[angle=-90]{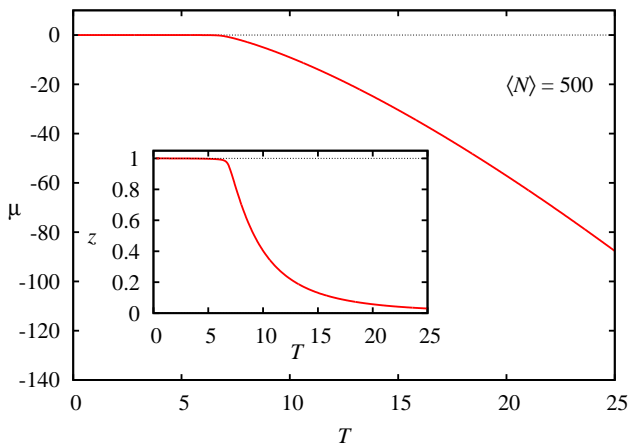}} 
\caption{The chemical potential $\mu$ and the fugacity $z$ as a function of temperature $T$ when $\varepsilon_0=0$ for a system of bosons in a three-dimensional HO with $\langle N\rangle = 500$. \todo[inline]{(Program: {\tt /home/vandijk/2012/zeros/Lee-Yang/december\_11/z\_mu\_vs\_T\_qp.f}).}}
\label{fig:02}
\end{figure}

 Quantum effects are manifest in a gas when the de Broglie thermal
wavelength of a  particle is larger than, or of the order of, the average 
interparticle spacing. 
BEC was first experimentally realized when neutral
$^{87}$Rb atoms\cite{anderson95} and $^{23}$Na atoms\cite{davis95} were magnetically trapped  in a HO potential at a few
  hundred degrees nano-Kelvin. 
For a large number of identical bosons, a sizable fraction of them
abruptly start occupying the lowest level even at a temperature much, 
much larger than the energy spacing $\hbar\omega$. This is the
condensation temperature. We follow the treatment of  Ketterle and
van Druten~\cite{ketterle96} to find the behaviour of the chemical potential $\mu$ as
the temperature is lowered. Here both the temperature and chemical
potential are taken to be real. Consider $N$ ideal bosons in the grand
canonical ensemble occupying a
discrete spectrum of single-particle states with energies  
$\varepsilon_n$ at temperature $T$. Its grand partition function may
be written as\cite{*[{}]  [{, page 199.}]  huang65a}
\beq
{\cal Z}(\beta, z)=\prod_n \left(1-z \exp (-\beta \varepsilon_n)\right)^{-1}
\label{raj}
\eeq
We then get 
\begin{equation}\label{eq:14}
\begin{split}
\langle N\rangle & =z \frac{\partial}{\partial z} \ln {\cal Z}
=\sum_{n=0}^\infty \dfrac{e^{\textstyle -\beta(\varepsilon_n-\mu)}}{1-e^{\textstyle -\beta(\varepsilon_n-\mu)}} \\
& = 
\sum_{n=0}^\infty \sum_{l=1}^\infty e^{\textstyle -\beta l\varepsilon_n}z^l =
\sum_{l=1}^\infty z^l Z_1(l\beta),
\end{split}
\end{equation}
where $Z_1(\beta)$ is the exact one-particle partition function 
and $z=e^{\textstyle\beta\mu}$. Since the occupancy factor
of a state has to be positive, it follows from the first term on the
RHS that the smallest value  $\mu$ can take is $\varepsilon_0$, the  
lowest energy single-particle state.  In the following, we choose 
$ \varepsilon_0=0$, so that $z\leq 1$, and the power series in $z$ does
not involve large numbers. Up till now the formulae are general. We now 
specialize to a shifted harmonic oscillator energy spectrum, which in one
dimension is given by $n\hbar\omega$, with $n$ going from 
zero to $\infty$.
For an isotropic three-dimensional harmonic oscillator $Z^{(3d)}_1(\beta) = \left(Z_1^{(1d)}(\beta)\right)^3$.  Thus we write for the three-dimensional HO 
\begin{equation}\label{eq:15}
\langle N\rangle  = \sum_{l=1}^\infty z^l \left(Z_1^{(1d)}(l\beta)\right)^3.
\end{equation}
Note that with the choice of zero-energy ground state,  $Z_1^{(1d)}(\beta)=
1/(1-\exp(-\beta\omega))$. In Fig.~\ref{fig:02}, we plot the variation
of the chemical potential and the fugacity as a function of $T$ for 
a system of trapped bosons in a three dimensional isotropic harmonic 
oscillator when $\langle N\rangle=500$ (we have set $\omega=1$).   
Note that the constraint on $\langle N\rangle$ makes $\mu$, and therefore $z$
temperature dependent.
There is no discontinuity in $\mu$ or $z$ for a finite number of
particles, but there is  a hint of rapid turning to a plateau in both
cases near $T=7$. This gets more  pronounced as $\langle N\rangle$ gets larger.
Similarly we can calculate the average energy in the grand canonical 
ensemble,
\beq\label{eq:16}
\langle E\rangle =\sum_n\frac{\varepsilon_n}{e^{\textstyle\beta(\varepsilon_n-\mu)}-1}
=-\sum_{l=1}^{\infty}\frac{z^l}{l}\frac{\partial }{\partial
\beta}Z_1(\beta l)
\eeq
For later use, we write the following expression for $\ln{\cal Z}(\beta,z)$ from Eq.~(\ref{raj}),
\beq
\ln {\cal Z}(\beta, z)=-\sum_n  \ln (1-z \exp(-\beta \varepsilon_n))
\label{eq:17}
\eeq
Noting that $\displaystyle \ln (1-x)=-\sum_{l=1}^{\infty}
\dfrac{x^l}{l}$, a few steps  
give 
\beq
\ln {\cal Z}(\beta, z)=\sum_{l=1}^{\infty} \frac{z^l}{l} Z_1(l\beta)~.
\label{eq:18}
\eeq
where 
\beq
Z_1(l\beta)=(1-\exp(-l\beta))^{-3}.
\label{eq:19}
\eeq
The  derivative of $\langle E\rangle$ with respect to $T$ gives the heat capacity. 
             
\begin{figure}[ht]
\centering
\resizebox{3.5in}{!}{\includegraphics[angle=-90]{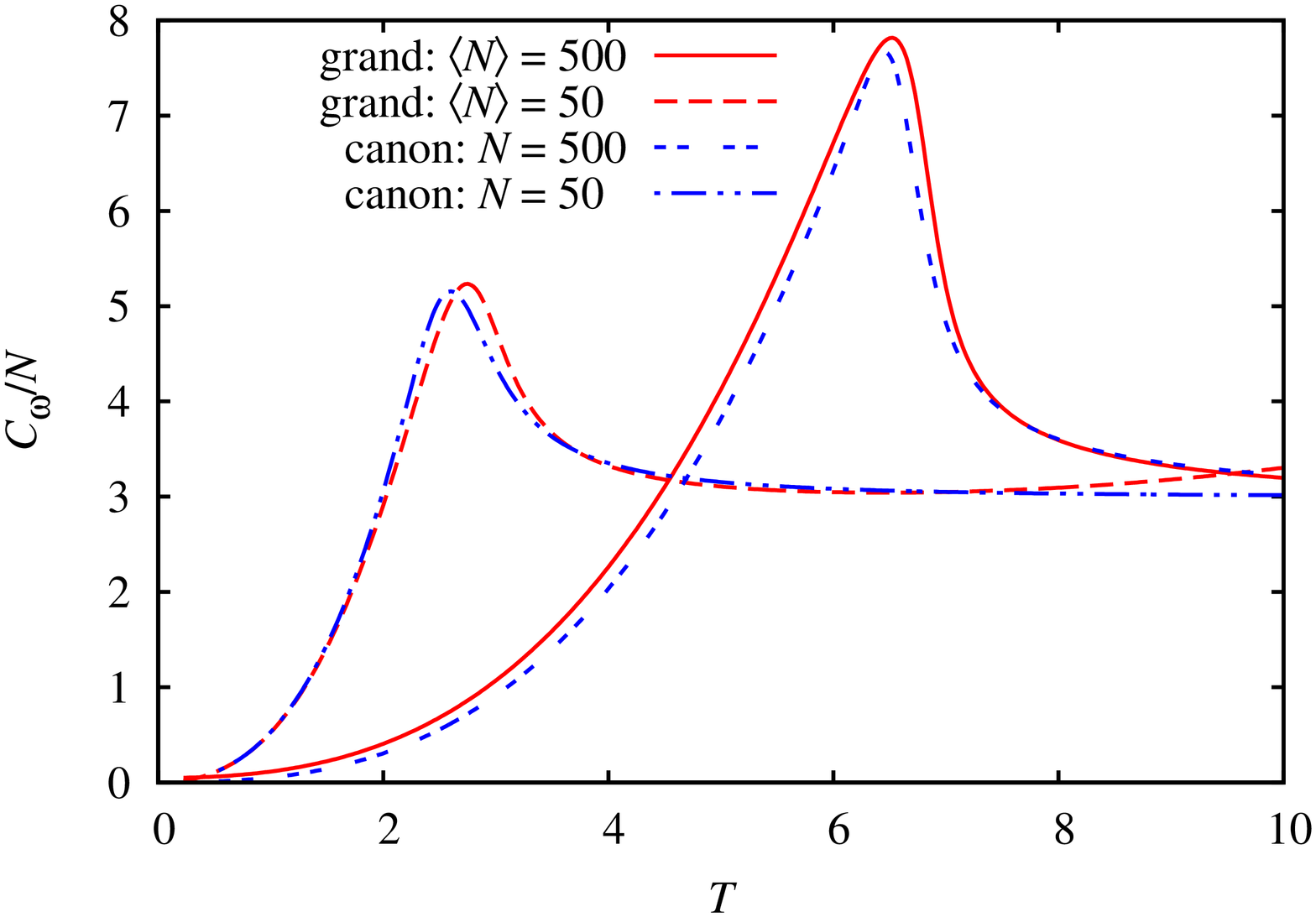}} 
\caption{The heat capacity per particle for a system
  of trapped bosons as a function of temperature.  The calculations based on the grand partition function, i.e., Eq.~(\ref{eq:16}), are labelled ``grand".  The graphs labelled ``canon" are based on the canonical partition function discussed later. \todo[inline]{The graph grand: $N=50$ seems to be correct.  One needs to explain why as $T$ becomes large the specific heat increases.  In my program, I started with $N=500$ and slowly decreased $N$.  As $N$ gets smaller to 100 say, then the specific heat per particle begins to increase as $T$ increases.  It is noticeable on the graph for $N=50$.}  }
\label{fig:03}
\end{figure}

 In Fig.~\ref{fig:03}, the heat capacity per
particle at constant $\omega$ is shown for $\langle N\rangle =500$ and $\langle N\rangle =50$ based on the grand partition analysis of this section. 
  Next we shall compare these results with the
canonical  formalism, which requires knowledge of $Z_N(\beta)$. 
 For finite $N$ and $\langle N\rangle$, the canonical and grand canonical ensembles  may yield  different results.
 
Before concluding this section, we note that for ideal bosons, in the
thermodynamic limit, there
is an analytical simple pole at $z=1$, the condensation point. This
may be seen from Eq.~(\ref{raj}), which shows that there are poles at 
\beq\label{eq:20a}
z_n=\exp(\beta \epsilon_n). 
\eeq
With our choice of $\epsilon_n \geq 0$, the RHS above is $\geq 1$. 
But as the inset of Fig.~\ref{fig:02} shows, physically allowed $z\leq 1$.  Therefore from Eq.~(\ref{eq:20a}), the only pole in the physical region is at $z=1$. 
\begin{figure*}[th]
\centering                                                                                      
$\begin{array}{c@{\hspace{0.0in}}c@{\hspace{0.0in}}}                                                                                                             
\resizebox{3.5in}{!}{\includegraphics[angle=-90]{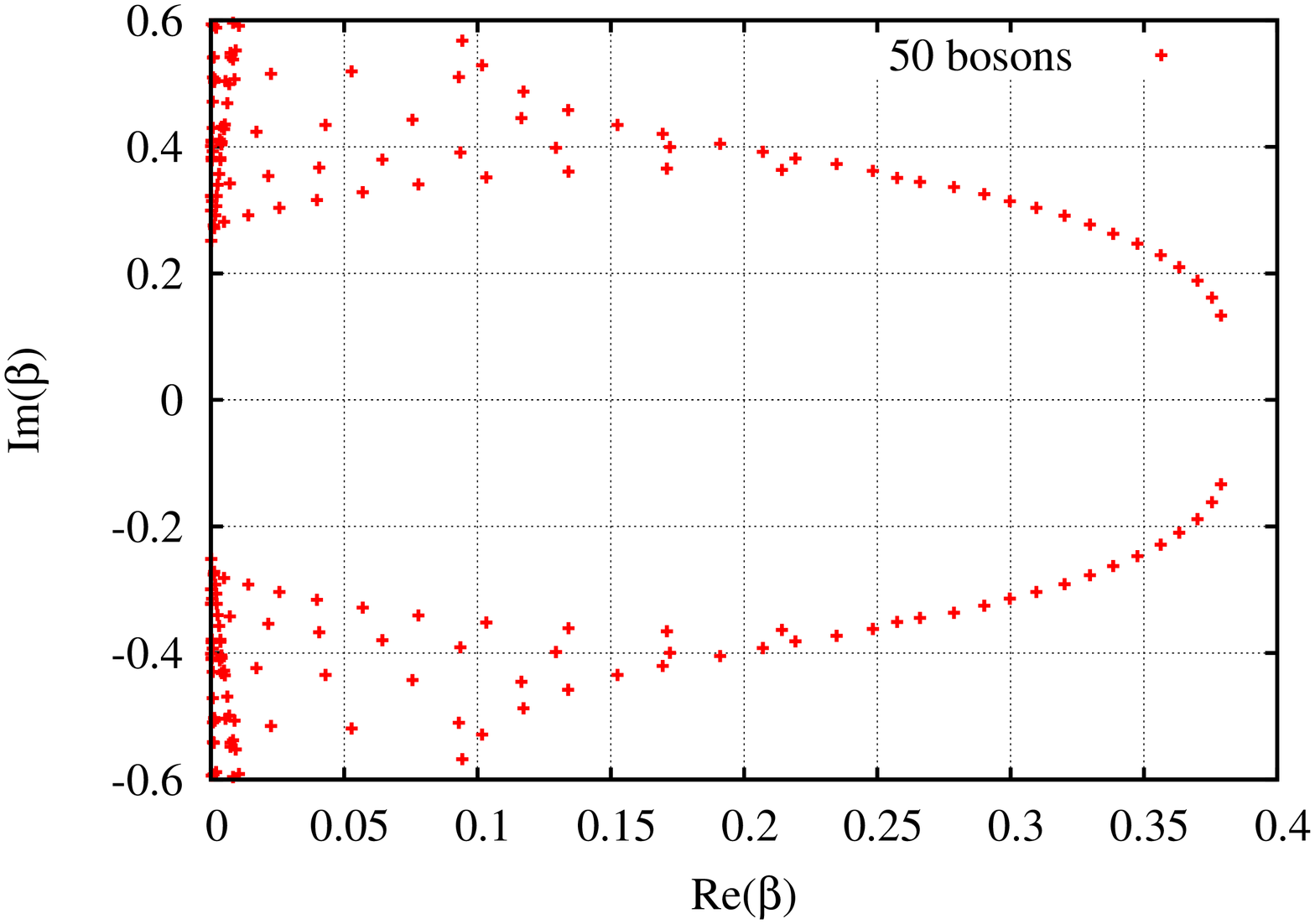}} &
\resizebox{3.5in}{!}{\includegraphics[angle=-90]{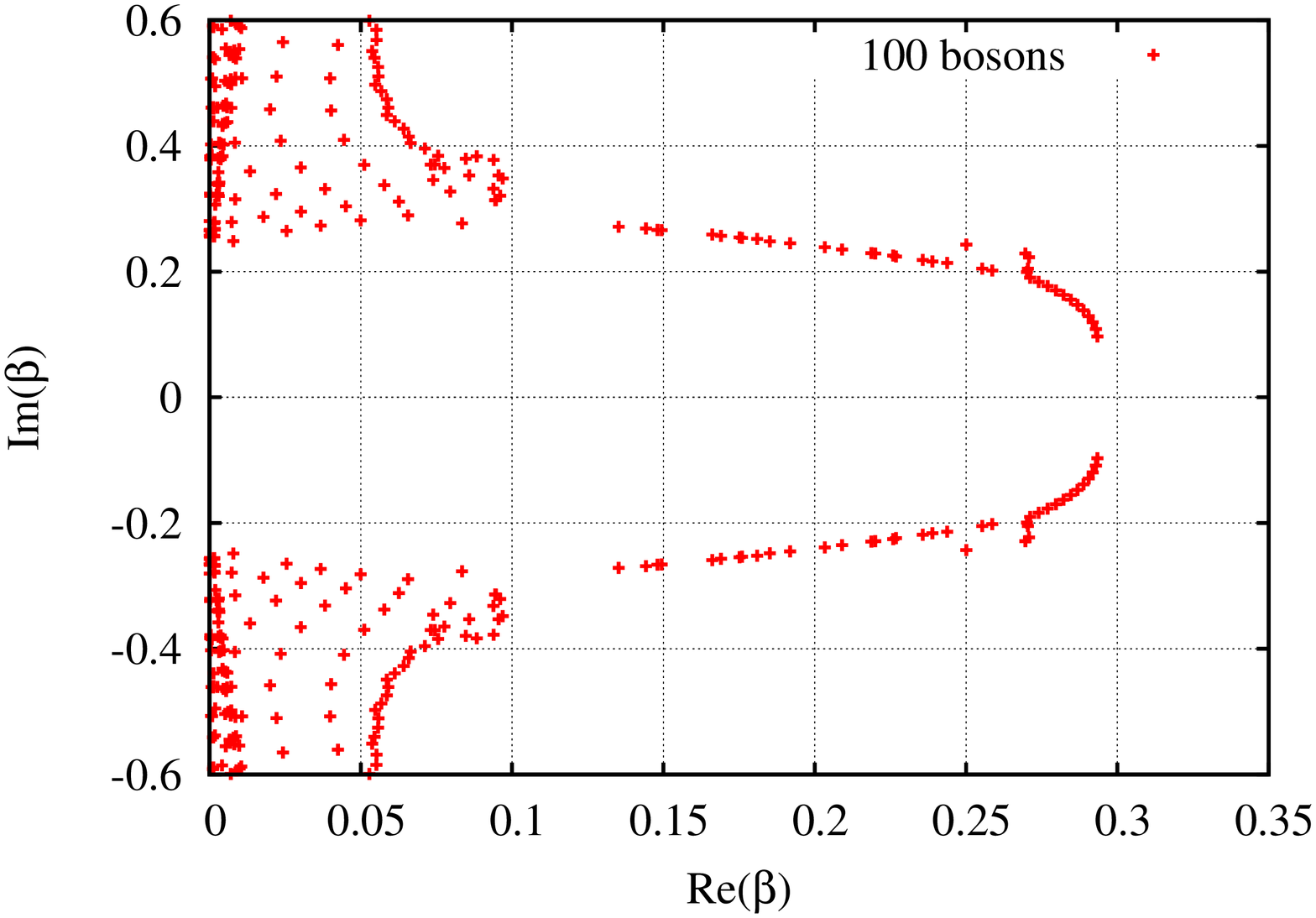}}
\end{array}$
\caption{The Fisher zeros for a systems of 50 or 100 trapped bosons taking into account the exact discrete energy spectrum. }
\label{fig:04}
\end{figure*} 

\subsection{Calculation of $ Z_N(\beta)$}
In order to obtain the grand partition function ${\cal Z}(\beta, z)$ from
Eq.~(\ref{eq:02}),  we need to calculate the canonical partition
function $Z_N(\beta)$. As we shall see soon, the zeros of the
canonical partition function on the complex $\beta$ plane (called the
Fisher zeros) are interesting in their own right. For an ideal boson
or fermion gas, $Z_N(\beta)$ may be obtained from $Z_1(\beta)$ using a
recursion relation.\cite{borrmann93}  We give an outline of the derivation of this
important relation for ideal bosons. 
 We start with the relation (\ref{eq:18}). 
On further expanding the exponential in a power series, and equating 
 power by power to the series  given by Eq.~(\ref{eq:02}),
\beq
{\cal Z}(\beta, z)=1+z Z_1(\beta)+z^2 Z_2(\beta)+z^3 Z_3(\beta)+\dots,
\label{eq:20}
\eeq 
the desired recursion relation emerges, which for bosons is given by~\cite{borrmann93} 
\beq
Z_N(\beta)=\frac{1}{N}\sum_{n=1}^NZ_1(n\beta) Z_{N-n}(\beta).
\label{eq:21}
\eeq
It is now straightforward to obtain
$\displaystyle \langle E\rangle =-\frac{\partial}{\partial\beta}\ln Z_N(\beta)$, and its
derivative with respect to $T$ to obtain the heat capacity.   In Fig.~\ref{fig:03} we plot $\displaystyle \dfrac{C_\omega}{N}=\dfrac{1}{N}\dfrac{\partial\langle E\rangle}{\partial T}$ as the graphs labelled ``canon".

\subsection{Fisher zeros for ideal bosons}

Fisher pointed out that there are complex zeros of the 
canonical partition function $Z_N(\beta)$ on the  
complex  $\beta$ plane. For a fixed $N$, the number of zeros on the 
complex plane, denoted by $\beta_r$, is finite. Therefore $Z_N(\beta) $ may be 
expressed as a finite product  $\prod{\left(1-\frac{\ts\beta}{\ts\beta_r}\right)}$. 
This is unlike the Lee-Yang zeros whose validity was contingent on 
short-range interparticle repulsion. 
At a phase transition, the complex Fisher zeros  
close in on the real $\beta$ axis. 
The Helmholtz free energy for an $N$-particle system is given by 
$\displaystyle F=-\dfrac{1}{\beta} \ln Z_N(\beta)$. 
Since $Z_N(\beta)$ is a sum of exponential positive terms in
$\beta$, and is larger than unity (the contribution of the state at
$\varepsilon =0$ ), $F$ does not change sign, and always remains negative. 
Nevertheless, $Z_N(\beta)$ is an entire function of $\beta$ in the
complex plane. For trapped bosons in a $3$-dimensional HO, we may use
the exact $Z_1(\beta)$ given by Eq.~(\ref{eq:19}). Then, 
using the recursion relation (\ref{eq:21}), $Z_N(\beta)$ is a
polynomial in the variable $y=e^{\textstyle -\beta}$ in this example of a
$3$-dimensional HO.\cite{schmidt99}  (See Appendix~\ref{appen:a}.)  The exact form of $Z_1(\beta)$ is taken, but 
whether $\varepsilon_0=3/2$ or $\varepsilon_0 =0$ the zeros occur at the same positions in the complex $\beta$ plane. Only a small
fraction of the zeros are shown in the Fig.~\ref{fig:04}. \todo[noline]{Obtain a similar graph when $\varepsilon_0=0$.  (Done.)}

\begin{figure}[!htb]
\centering
\resizebox{3.5in}{!}{\includegraphics[angle=-90]{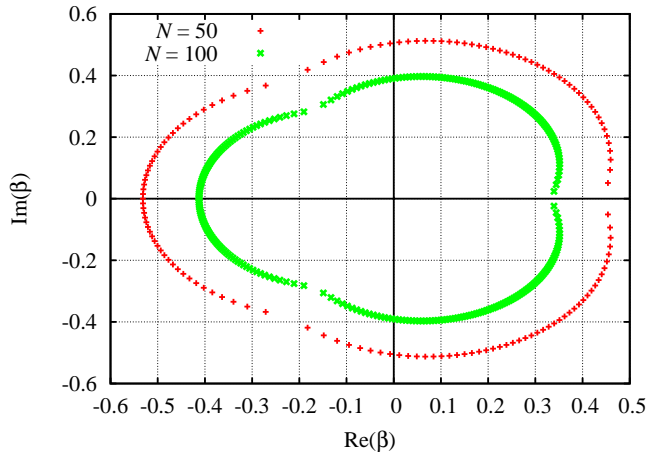}} 
\caption{The Fisher zeros for systems of 50 or 100 trapped bosons with a continuous energy distribution.}
\label{fig:05a}
\end{figure}

Even for $N=50$, there is a tendency for the zeros to approach
the real $\beta$-axis, but clearly the system is not condensed. We
also display the plot of zeros for $N=100$ ideal bosons.   Calculations
are much easier if one uses $Z_1(\beta)=1/\beta^3$, which is the
leading term of the exact Eq.~(\ref{eq:19}). This corresponds to a 
continuous single-particle density of states that grows quadratically.
In Fig.~\ref{fig:05a}, we show the pattern of complex zeros of the corresponding
$Z_N(\beta)$ for $N=50$ and $N=100$. Comparison with Fig.~\ref{fig:04} shows 
considerable difference:  the number of zeros being much smaller for
the case of continuous density of states. In the latter, $\beta_c$ is
about $10$ percent higher. In both, the estimated condensation temperaure   
$T_c$ increases approximately as $N^{1/3}$.

\section{Concluding Remarks}

We have shown that the advent of a phase transition in a system is
reflected in the pattern of the complex zeros of the partition
function. Strictly speaking, a phase transition takes place only in
the thermodynamic limit. But even for a finite system with relatively 
small number of particles, the pattern of complex zeros begin to close
in on the real fugacity or inverse temperature axis.  For the grand
partition function Lee-Yang zeros, it is imperative to have short-range repulsion in the interparticle interaction, whereas for the
Fisher zeros of the canonical partition function, this is not
necessary. In the case of BEC, a signal of a phase transition is a 
peak in the heat capacity per particle on the real temperature axis,
as shown in Fig. 3.  This peak shows up nicely even for $\langle N\rangle =50$. In 
Fig. 4, the complex Fisher zeros for $N=50$ appear to close in on the
real axis at the same temperature. 
 The Lee-Yang zeros in the van der Waal gas seem to close in towards 
the real z axis for $T\leq T_c$.
Finally, we note that even though there are no Lee-Yang zeros for the
ideal Bose gas, the grand canonical partition function has a simple
pole at $z=1$, which was already noticed by Kastura.\cite{katsura63} 

\acknowledgments
The authors are grateful to Professor Akira Suzuki for helpful discussions.

\vspace{1cm}

\appendix

\section{Fisher zeros} 
\label{appen:a}
The Fisher zeros are the zeros of the $N$-particle canonical partition function $Z_N(\beta)$ in the complex $\beta$ plane.  Given $Z_0(\beta)=1$ and $Z_1(\beta)=\sum_{n=0}^\infty e^{\ts -\beta E_n}$, we can obtain $Z_N(\beta)$  by the recursion\cite{borrmann93} 
\begin{equation}\label{eq:a01}
Z_N(\beta)=\dfrac{1}{N}\sum_{k=1}^NZ_1(k\beta)Z_{N-k}(\beta).  
\end{equation}
If $Z_1(\beta)$ is the canonical partition function of a single particle in a three-dimensional harmonic oscillator well, then 
\begin{equation}\label{eq:a03}
 Z_1(\beta)=\left(\sum_{n=0}^\infty e^{\displaystyle -\beta\varepsilon_n}\right)^3 =\left(\dfrac{e^{\ts -\beta\hbar\omega/2}}{1-e^{\ts -\beta\hbar\omega}}\right)^3.
\end{equation} 
where $\varepsilon_n=(n + 1/2)\hbar\omega$.
The calculation of $Z_N(\beta)$ can be simplified by introducing~\cite{schmidt99} $y = e^{\ts -\beta\hbar\omega}$ so that
\begin{equation}\label{eq:a04}
Z_N(y) = \dfrac{y^{3N/2}}{\prod_{j=1}^N(1-y^j)^3}P_N(y).
\end{equation}
The recursion relation~(\ref{eq:a01}) is reformulated as $P_0(y)=P_1(y)=1$ and
\begin{equation}\label{eq:a05}
P_N(y)  = 
\dfrac{1}{N}\sum_{k=1}^N  \dfrac{\prod_{j=N-k+1}^N(1-y^j)^3}{(1-y^k)^3} P_{N-k}(y).
\end{equation}
The $P_N(y)$ is a polynomial in $y$ and when it is zero so is $Z_N(y)$.   
In the case that $\varepsilon_n=n\hbar\omega$ rather than $(n+1/2)\hbar\omega$ we have
\begin{equation}\label{eq:a06}
Z_1(y)= \dfrac{1}{(1-y)^3}, \ \ \ Z_N(y)=\dfrac{1}{\prod_{j=1}^N(1-y^j)^3} P_N(y),
\end{equation}
where the $P_N$ satisfy the same recursion~(\ref{eq:a05}).  Thus the Fisher zeros will be the same irrespective of $\varepsilon_0 = 0$ or $\hbar\omega/2$.

Since $P_N(y)$ is a polynomial the number of zeros is equal to its degree which increases rapidly with particle number.  For example, for $N=50$ there are 3495 zeros.  We determine a subset to give a clear indication that the pattern of zeros pinches the positive real $\beta$ axis.

In order to numerically obtain the zeros we use the Laguerre method since this method ``is guaranteed to converge to a (zero) from any starting point."~\cite{*[{}] [{, pages 263ff.}] press_F86}  Once a zero is found we deflate the polynomial to obtain a second distinct zero and repeat.  For larger $N$ the zeros may be close together so we check the accuracy of the zero $y_0$ by considering a small circle centred on $y_0$ in the complex $y$ plane and by ensuring that the curves $\mathrm{Re}(P_N(y))=0$ and $\mathrm{Im}(P_N(y))=0$ intersect inside the small disc. By using a radius of say $10^{-5}$  we have an estimate of the precision of the zero.  Since we are interested in zeros that pinch the positive real $\beta$ axis, we limit the variable $y$ so that $|y|<1$ which results in $\mathrm{Re}(\beta)>0$.    

%

\end{document}